\documentclass[english]{llncs}

\usepackage{graphicx}
\usepackage{mathpartir}
\usepackage{xspace}
\usepackage{proof}
\usepackage{amsfonts}
\usepackage{amsmath}
\usepackage{stmaryrd}
\usepackage{verbatim}
\usepackage{paralist}
\usepackage{float}
\usepackage{listings}
   \usepackage{url}
   \usepackage[colorlinks=true,urlcolor=blue,citecolor=blue,linkcolor=blue]{hyperref}
\usepackage{caption}
\usepackage{subcaption}
\usepackage[outercaption]{sidecap}
\usepackage{wasysym}

\usepackage[T1]{fontenc}
\usepackage{libertine}
\usepackage[scaled=0.8]{beramono}

\usepackage{microtype}

\makeatletter
\renewcommand{\paragraph}{%
  \@startsection {paragraph}{4}%
  {\z@ }{-6\p@ \@plus -4\p@ \@minus -4\p@ }{-0.5em \@plus -0.22em \@minus -0.1em}%
  {\normalfont \normalsize \itshape }%
}
\makeatother

\usepackage{xparse}
\ExplSyntaxOn
\NewDocumentCommand{\xsubfigure}{ m m }
 {% #1 is a symbolic key, #2 is a list of key-value pairs
  \roly_xsubfigure:nn { #1 } { #2 }
 }
\NewDocumentCommand{\makerow}{ m }
 {% #1 is a list of symbolic keys
  \roly_makerow:n { #1 }
 }

\keys_define:nn { roly/subfigures }
 {
  width .tl_set:N = \l_roly_subfig_width_tl,
  body .tl_set:N = \l_roly_subfig_body_tl,
  caption .tl_set:N = \l_roly_subfig_caption_tl,
 }

\dim_new:N \l_roly_row_height_dim
\box_new:N \l_roly_body_box

\cs_new_protected:Npn \roly_xsubfigure:nn #1 #2
 {
  \prop_if_exist:cTF { l_roly_subfig_#1_prop }
   {
    \prop_clear:c { l_roly_subfig_#1_prop }
   }
   {
    \prop_new:c { l_roly_subfig_#1_prop }
   }
  \keys_set:nn { roly/subfigures } { #2 }
  \prop_put:cnV { l_roly_subfig_#1_prop } { width } \l_roly_subfig_width_tl
  \prop_put:cnV { l_roly_subfig_#1_prop } { body } \l_roly_subfig_body_tl
  \prop_put:cnV { l_roly_subfig_#1_prop } { caption } \l_roly_subfig_caption_tl
 }

\cs_new_protected:Npn \roly_makerow:n #1
 {
  \dim_zero:N \l_roly_row_height_dim
  \clist_map_inline:nn { #1 }
   {
    \hbox_set:Nn \l_roly_body_box
     {
      \prop_get:cn { l_roly_subfig_##1_prop } { body }
     }
    \dim_compare:nT { \box_ht:N \l_roly_body_box > \l_roly_row_height_dim }
     {
      \dim_set:Nn \l_roly_row_height_dim { \box_ht:N \l_roly_body_box }
     }
   }
  \clist_map_inline:nn { #1 }
   {
    \begin{subfigure}[t]{ \prop_get:cn { l_roly_subfig_##1_prop } { width } }
    \raggedright
    \vspace{0pt} % for top alignment
    \parbox[t][\l_roly_row_height_dim]{\textwidth}{
      \prop_get:cn { l_roly_subfig_##1_prop } { body }
    }
    \caption{ \prop_get:cn { l_roly_subfig_##1_prop } { caption } }
    \end{subfigure}
    \hspace{2em} % some space between the objects in a row
   }
   \unskip\\ % end up the row
 }
\ExplSyntaxOff

\DeclareFontFamily{U}{matha}{\hyphenchar\font45}
\DeclareFontShape{U}{matha}{m}{n}{
      <5> <6> <7> <8> <9> <10> gen * matha
      <10.95> matha10 <12> <14.4> <17.28> <20.74> <24.88> matha12
      }{}
\DeclareSymbolFont{matha}{U}{matha}{m}{n}

\DeclareMathSymbol{\second}{3}{matha}{"32}
\DeclareMathSymbol{\third}{3}{matha}{"33}
\DeclareMathSymbol{\fourth}{3}{matha}{"34}

\newcommand{\etal}{et al.\xspace}

\newcommand{\figref}[1]{Fig.~\ref{fig:#1}}

\newcommand{\figsubref}[2]{Fig.~\ref{fig:#1}{#2}}
\newcommand{\figsubreftwo}[3]{Figs.~\ref{fig:#1}{#2} and \ref{fig:#1}{#3}}

\newcommand{\secref}[1]{\S\ref{sec:#1}}

\newcommand{\tableref}[1]{Table~\ref{table:#1}}

\newcommand{\ProPub}{\textsf{ProPub}\xspace}
\newcommand{\PROV}{\textsf{PROV}\xspace}
\newcommand{\ProvAbs}{\textsf{ProvAbs}\xspace}
\newcommand{\Zoom}{\textsf{ZOOM}\xspace}

\newcommand{\define}[1]{\emph{#1}}

\newcommand{\FULL}{\CIRCLE}
\newcommand{\NONE}{\Circle}
\newcommand{\HALF}{\LEFTcircle}

\newcommand{\featIntegrity}{Int\xspace}
\newcommand{\featDependency}{Dep\xspace}
\newcommand{\featAccess}{Acc\xspace}
\newcommand{\featQuery}{Qry\xspace}
\newcommand{\featSemantic}{Sem\xspace}
\newcommand{\featFormal}{Form\xspace}
\newcommand{\featConflict}{Conf\xspace}
\newcommand{\featMeta}{Meta\xspace}

\begin{document}
\author{James Cheney \and Roly Perera}
\institute{
\email{\textsf{\{jcheney, rperera\}@inf.ed.ac.uk}} \\
School of Informatics, University of Edinburgh
}

\title{An Analytical Survey of Provenance Sanitization }
\maketitle

\begin{abstract}
  Security is likely becoming a critical factor in the future adoption of
  provenance technology, because of the risk of inadvertent disclosure
  of sensitive information. In this survey paper we review the state of
  the art in secure provenance, considering mechanisms for controlling
  access, and the extent to which these mechanisms preserve provenance
  integrity. We examine seven systems or approaches, comparing features
  and identifying areas for future work.
\end{abstract}

\section{Introduction}

Automatically associating data with metadata describing its provenance
has emerged as an important requirement in databases, scientific
computing, and other domains that place a premium on reproducibility,
accountability or trust~\cite{moreau10ftws}. Providing such metadata
typically involves instrumenting a system with monitoring or logging
that tracks how results depend on inputs and on other, perhaps
untrustworthy, sources.

Publishing the entire provenance record associated with a computation is
not always feasible or desirable. Disclosing certain information may
violate security, privacy, or need-to-know policies, or expose sensitive
intellectual property. Sometimes the complete provenance record may be
too detailed for the intended audience, or may leak irrelevant
implementation detail. But simply omitting some of the provenance
information may leave it unable to certify the origins of the data
product.

We refer to the general problem of ensuring that provenance solutions
satisfy not only disclosure requirements but also security or privacy
requirements as the problem of \emph{provenance sanitization} or
\emph{provenance abstraction}. A number of approaches to provenance
sanitization have been proposed
recently~\cite{blaustein11pvldb,chebotko08,davidson11pods,davidson13icdt,dey11ssdbm},
sometimes under other names such as \emph{provenance views} or
\emph{provenance redaction}. These techniques have been developed mainly
for scientific workflow systems, where provenance is viewed as a
directed acyclic graph, as in the Open Provenance Model~\cite{opm11}.

Existing approaches have several elements in common. Typically, an
\emph{obfuscation policy} specifies the aspects of the provenance which
are to be hidden. A \emph{disclosure policy} may additionally specify
that certain other aspects of the provenance are to remain visible.
\emph{Sanitization} then involves transforming the provenance graph to
obtain a view which satisfies both the obfuscation and the disclosure
policies.

Few of the existing systems have been formally studied, and the security
guarantees they actually provide are unclear. Some do provide formal
guarantees, but are narrow in applicability or have other shortcomings.
Moreover, many systems provide some form of security or confidentiality
without considering the impact on the causal or explanatory role of
provenance. In this paper we review the state of the art in provenance
sanitization by reviewing seven systems or approaches:
\Zoom~\cite{biton08icde,cohen-boulakia08}, security
views~\cite{chebotko08}, surrogates~\cite{blaustein11pvldb},
\ProPub~\cite{dey11ssdbm}, provenance
views~\cite{davidson11pods,davidson13icdt}, provenance
abstraction~\cite{missier13}, and provenance
redaction~\cite{cadenhead11}.

\section{Related work}
\label{sec:related-work}

The relationship between security and provenance has been considered in
several survey or vision
papers~\cite{hasan07sss,braun08hotsec,lyle10tapp,martin12tapp}. This
paper focuses narrowly on provenance sanitization via graph
transformations; here we briefly mention some related topics.

\paragraph{Formal foundations.}

Chong~\cite{chong09tapp} gave an early definition of provenance-related
security policies. Cheney~\cite{cheney11csf} subsequently generalized
this approach to notions of \emph{disclosure} and \emph{obfuscation}
with respect to a query $Q$ on the underlying provenance, and a view $P$
of the provenance. Obfuscation is similar to (non-quantitative)
\emph{opacity} in computer security~\cite{bailliage04}, and means that
$P$ does not allow the user to determine whether the underlying
provenance satisfies $Q$. Disclosure means that $P$ preserves
$Q$-equivalence.

\paragraph{Secure provenance for evolving data.}

Provenance tracking is an especially critical issue for data that
changes over time~\cite{buneman06sigmod}, for which provenance can be
hard to recover after the fact. Work in this area to date includes
tamper-resistant provenance for databases~\cite{zhang09sdm}, use of
cryptographic techniques to ensure integrity of document version
history~\cite{hasan09tos}, and database audit log
sanitization~\cite{lu13vldbj}.

\section{Background concepts and terminology}
\label{sec:overview}

The solutions surveyed in \secref{survey} mainly target scientific
workflow systems, with similar notions of provenance; we review some
common concepts here. Some acquaintance with basic graph theory will be
useful. For more background on scientific workflow provenance, we refer
the reader to Davidson and Freire \cite{davidson08icmd}.

\paragraph{Workflow systems and provenance graphs.}

\begin{figure}[H]
\begin{center}
\includegraphics[scale=0.8]{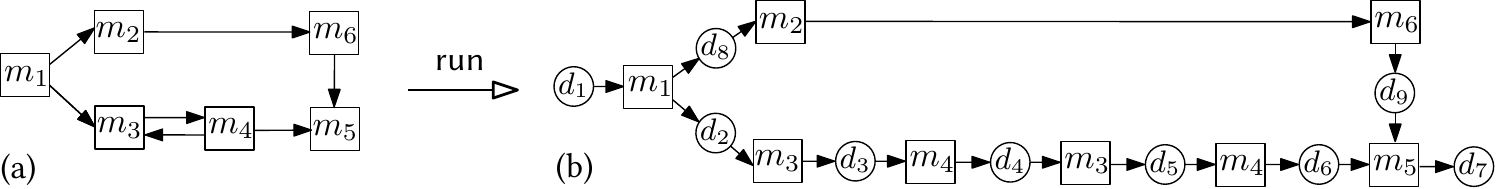}
\end{center}
\caption{Cyclic workflow, plus bipartite provenance graph for a possible run}
\label{fig:overview:workflow}
\end{figure}

A \define{workflow system}, or simply \define{workflow}, is a directed
graph capturing the high-level structure of a software-based business
process or scientific process. Nodes represent software components
called \define{modules}, or \define{tasks}. Edges represent links, or
\define{data channels}, connecting modules. Sometimes modules are
considered to have input and output \emph{ports} to which data channels
are connected. \figsubref{overview:workflow}{a} shows a simple workflow
with modules $m_1$ to $m_6$.

Node labels are typically used to identify modules. Iterative processes
can be modelled by cycles, if permitted, or via a built-in construct for
iteration. Workflow systems often support other coordination patterns
such as conditional branching and synchronisation which are beyond scope
of the systems considered here. Some permit \define{composite} modules,
i.e.~modules that contain other modules.

A \define{provenance graph} is a directed, acyclic graph (DAG) recording
the causal history of a data product. Often such a graph represents the
(coarse-grained) \emph{execution} of a software system, such as a
workflow; more generally, provenance graphs can describe ad hoc
processes or collaborations involving both human and software
components. The nodes of the graph represent participants, actions and
intermediate artifacts.

\figsubref{overview:workflow}{b} shows a provenance graph that captures
one possible execution of the workflow in
\figsubref{overview:workflow}{a}. The rectangular nodes, or
\define{activities}, represent invocations of modules; the circular
nodes $d_1$ to $d_9$, sometimes called \emph{entities}, record data
values passed between modules. Moreover activities yield entities, and
entities feed into activities; a graph that is partitioned in this way
is called \emph{bipartite}. Bipartiteness is just one of many possible
design choices for graph-structured provenance; for example, one could
add $d_1, \ldots, d_9$ as labels to the edges instead of using special
nodes.

When a provenance graph represents a run of an iterative process, each
module invocation must give rise to a distinct node, to maintain
acyclicity. If necessary additional tags on the node label can be used
to distinguish invocations of the same module.

\paragraph{Sanitizing provenance graphs}

The goal of provenance sanitization is to derive a \define{sanitized
  view} which hides or abstracts sensitive details of a provenance
graph, whilst preserving some of its disclosure properties. Typically
one wants the view itself to be a well-formed provenance graph.
\figref{overview:provenance-view} below illustrates a simple provenance
graph with two examples of views. On the right, tasks $c_1$ and $c_2$
have been abstracted into a single task $c_3$; on the left, entities
$d_2$ and $d_4$ and intermediate task $c_2$ have been abstracted into a
single entity $d_5$.

\begin{figure}[H]
\begin{center}
\includegraphics[scale=0.8]{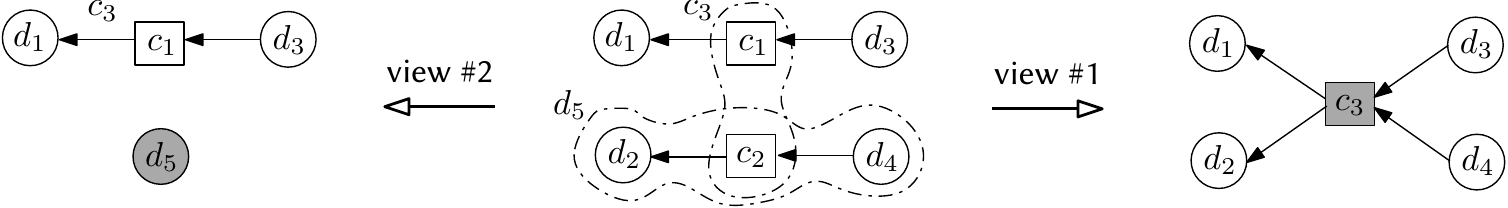}
\end{center}
\caption{Two possible views of a provenance graph}
\label{fig:overview:provenance-view}
\end{figure}

\noindent Both views are examples of \emph{quotients}, arguably the
simplest notion of graph view. One specifies a quotient of a graph $G =
(V,E)$ by giving a partitioning $V' = \{V_1, \ldots, V_n\}$ of its
nodes. The partitioning induces another graph $G' = (V',E')$ where there
is an edge $(V_i, V_j) \in E'$ iff there is an edge in $G$ between a
node of $V_i$ and a node of $V_j$, for any $i \neq j$. In
\figref{overview:provenance-view} the dotted border labeled $c_3$
determines a partitioning if we consider each of the remaining nodes to
inhabit a singleton partition; the dotted border labeled $d_5$
determines a different partitioning, under a similar assumption.

Quotients are a natural forms of provenance view as they preserve
\emph{paths}, which represent relationships of direct or indirect
dependency between nodes. If paths are preserved then related nodes are
mapped to related nodes in the view; in other words, every dependency in
the original graph gives rise to a dependency between the corresponding
view nodes. Quotients preserve paths but not edges; for example the
edges $(d_4, c_2)$ and $(c_2, d_2)$ have no counterpart in view \#2
because all three nodes are mapped to $d_5$. Indeed edge-preservation,
or \emph{homomorphism}, is a stronger property than we usually require
for provenance sanitization, where dependency is assumed to be reflexive
and transitive.

It can also be important to consider whether paths are \emph{reflected}:
whether nodes are related in the view \emph{only if} there exist related
nodes in the original graph which map to those nodes in the view. This
too can be understood in terms of dependency, since it means that every
reported dependency arises from a dependency between corresponding nodes
in the original graph. Quotients do not in general reflect paths,
because they coarsen the dependency relation: in view \#1, for example,
$d_1$ now appears to depend on $d_4$, and $d_2$ on $d_3$. This can be
problematic if it violates cardinality constraints, such as a
requirement that every artifact be generated by at most one activity
\cite{moreau13wc3}.

\section{Survey of techniques for provenance sanitization}
\label{sec:survey}

In the \Zoom system of \textbf{Biton, Cohen-Boulakia and Davidson}
\cite{biton08icde,cohen-boulakia08}, the user obtains a provenance view
by first defining an abstract workflow view. A \Zoom workflow is a
directed graph of atomic modules; a provenance graph is a DAG of
invocations with edges labeled with runtime values. A workflow view is a
quotienting which partitions the system into composite modules; for a
given run of the workflow, the corresponding ``quotient run'' can then
be obtained automatically by deriving invocations of each composite
module from the invocations of its constituent modules.

\begin{figure}[H]
\begin{center}
\includegraphics[scale=0.8]{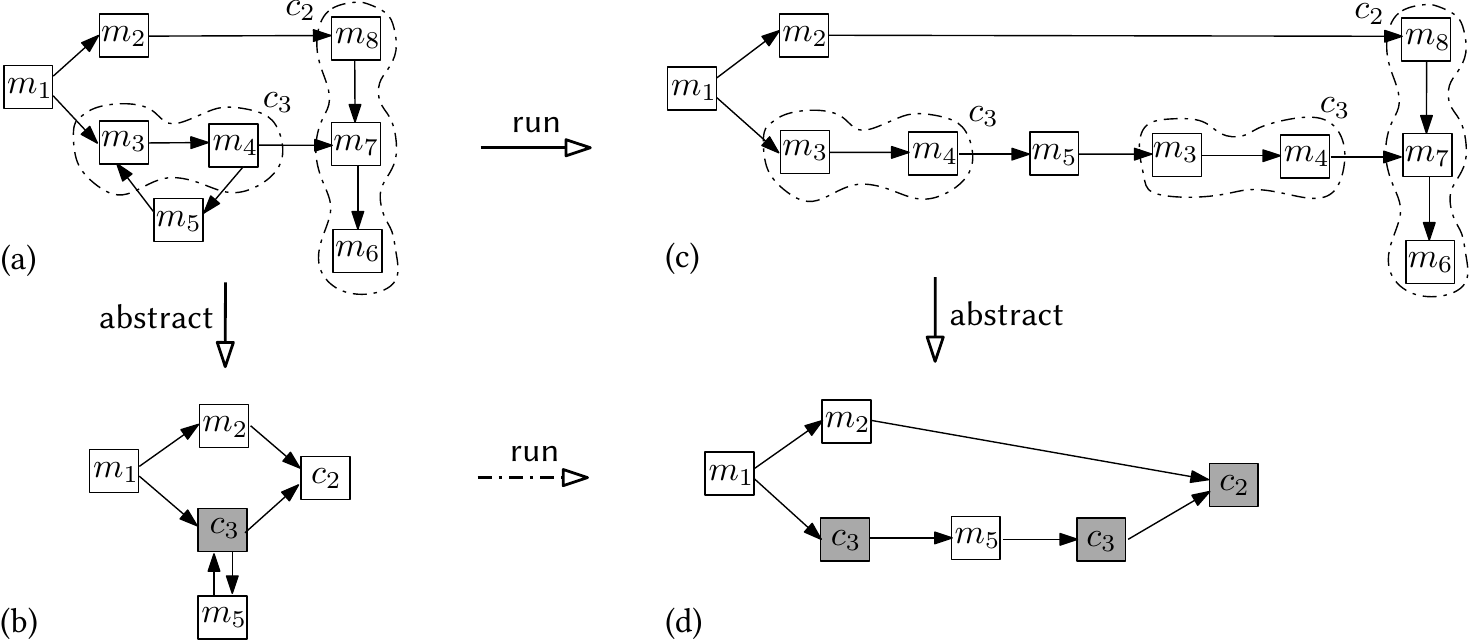}
\end{center}
\caption{\Zoom: deriving a provenance view from a workflow view}
\label{fig:survey:biton}
\end{figure}

\figref{survey:biton} illustrates the \Zoom approach. In
\figsubref{survey:biton}{a} we see the original workflow with the
partitioning identified by dashed borders labeled $c_2$ and $c_3$. The
modules $m_1$, $m_2$ and $m_5$ are assumed to be in singleton
partitions. The induced workflow view is shown in
\figsubref{survey:biton}{b}. Then, \figsubref{survey:biton}{c} shows an
execution of the workflow with data labels omitted; here the dashed
borders represent a partitioning of the \emph{invocations} corresponding
to invocations of the composite modules $c_2$ and $c_3$.
\figsubref{survey:biton}{d} shows the the corresponding quotient run
where each node is mapped to its equivalence class.

\Zoom is not overtly motivated by security, but its views can be seen as
abstracting away uninteresting parts of the graph while ensuring
user-identified ``relevant'' parts remain visible.
\Zoom is unique in respecting the semantic relationship between program
and provenance, as alluded to by the dotted \textsf{run} arrow relating
\figsubreftwo{survey:biton}{b}{d}. Moreover being able to derive
provenance views from \emph{ex post facto} modularisations of a workflow
is extremely powerful. However, it seems unlikely that their method for
doing so (sketched only briefly in the papers) will generalise to
workflows with non-trivial control flow or settings where submodules are
shared by composite modules. In \cite{cohen-boulakia08}, most of the
focus is on workflow views instead, in particular a method for deriving
workflow views that preserve and reflect certain structural properties
of the workflow, given a user-specified set of modules that are of
interest.

The \textbf{security views} of \textbf{Chebotko, Chang, Lu, Fotouhi and
  Yang} \cite{chebotko08} provide both access control and abstraction
for scientific workflow provenance. Their workflows are DAGs with
additional structure to model hierarchical tasks; the data channels of a
composite task are those of its constituent tasks that cross the
boundary of the composite task, relating composite tasks to the
partitions of a quotient view. However, composite tasks are fixed
features of the workflow rather than on-the-fly abstractions as in
\Zoom, above. Being acyclic, workflows are unable to represent
iteration.

To obtain a security view, one first specifies the accessibility of the
various tasks and data channels, marking each element as accessible or
inaccessible. Inheritance rules define the accessibility of an element
if it is not given explicitly. Access control can be specified down to
the level of individual ports; consistency constraints ensure that (for
example) a data value inaccessible on one port is not accessible via
another port. The access specification is then used to derive a
provenance view from which inaccessible data values, tasks and channels
have been removed.

\begin{figure}[H]
\begin{center}
\includegraphics[scale=0.8]{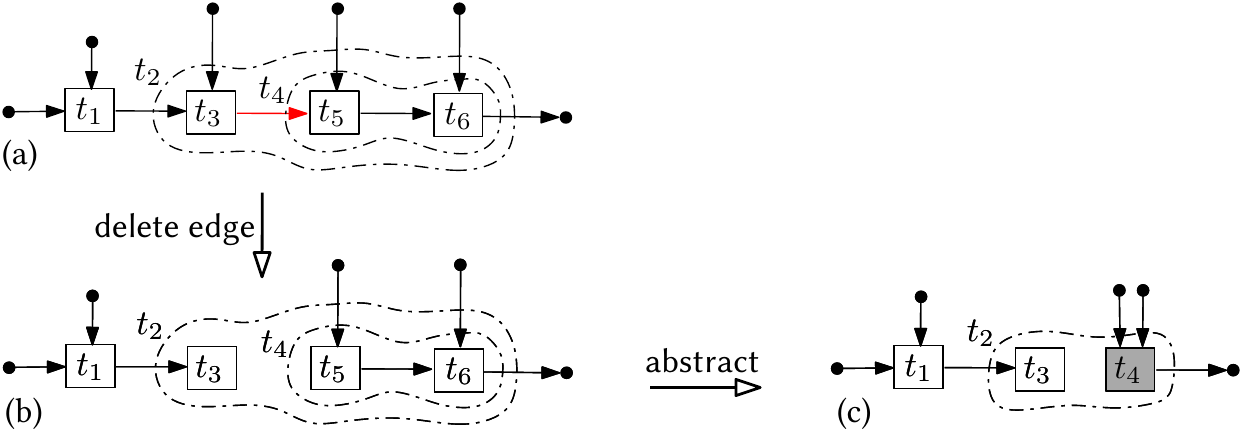}
\end{center}
\caption{Security views: combining abstraction with access control}
\label{fig:survey:chebotko}
\end{figure}

\figsubref{survey:chebotko}{a} shows a run of a hierarchical workflow
with two levels of composite task; both data nodes and ports have been
elided for brevity. A node written as $\bullet$ indicates an input or
output. In \figsubref{survey:chebotko}{b}, the data channel between
$t_3$ and $t_5$ has been deleted to conform to the access specification.
Although dummy nodes, similar to the \emph{surrogates} of Blaustein
\etal below \cite{blaustein11pvldb}, may be added to the view to
preserve well-formedness constraints,
% properties such as bipartiteness,
more general
integrity requirements are not considered. For example once the edge
between $t_3$ and $t_5$ has been deleted, the view no longer preserves
dependencies, and so its ability to provide a full account of the output
is compromised. Access control can however be combined with quotienting.
In \figsubref{survey:chebotko}{c} the composite module $t_4$ has been
abstracted to a single node with two inputs, preserving the dependency
structure of \figsubref{survey:chebotko}{b}, even though the latter
view is
unsound.

\textbf{Blaustein, Chapman, Seligman, Allen and Rosenthal}
\cite{blaustein11pvldb} present an approach based on \emph{surrogates}.
They define a \emph{protected account} of a graph $G$ to be any graph
$G'$, along with a path-preserving function from the nodes of $G'$ to
the nodes of $G$. Since by definition every path in the view has an
image in the original graph, a protected account necessarily reflects
dependencies, but in general does not preserve them. Surrogates are a
mechanism for publishing dependency information in a way that still
protects sensitive nodes and edges.

\begin{figure}[H]
\begin{center}
\includegraphics[scale=0.8]{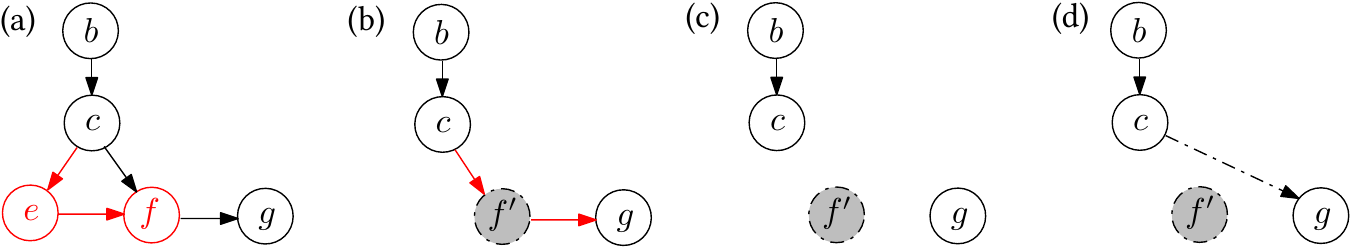}
\end{center}
\caption{Surrogates: provenance graph, plus three protected accounts}
\label{fig:survey:protected-accounts}
\end{figure}

\figsubref{survey:protected-accounts}{a}, adapted from
\cite{blaustein11pvldb}, shows a typical graph with sensitive nodes and
edges in red. \figsubref{survey:protected-accounts}{b} shows a protected
account where $e$ has been deleted and $f$ replaced by a surrogate $f'$,
shown with a dotted border, that hides its sensitive data (perhaps its
identity). The view in \figsubref{survey:protected-accounts}{c} hides
two more edges, breaking the indirect dependency between $c$ and $g$.
This is repaired in \figsubref{survey:protected-accounts}{d} by a
surrogate edge (dotted arrow).

Blaustein \etal's approach has three components: user privileges, which
allow the graph provider to control graph access down to the level of
individual ports; an algorithm for protecting graphs by deleting nodes
and edges and adding surrogates; and metrics for analyzing disclosure
and obfuscation properties of the resulting graph. For a given set of
user privileges, their algorithm purportedly obtains a protected account
which is ``maximally informative'', according to a utility metric
derived from the proportion of $G$-paths retained in $G'$ plus the
similarity of each node in $G'$ to its counterpart in $G$. However
definitions given are rather informal, and the theorems lack proofs,
making this claim hard to evaluate.

Even when a protected account satisfies a particular obfuscation policy,
an attacker may still be able to infer the original graph $G$ from $G'$.
To study this, Blaustein \etal introduce the notion of \emph{opacity}, a
measure of the difficulty of inferring an edge in $G$ that is not
present in $G'$, given a user-supplied model of the attacker. (The
notion of opacity in the security literature \cite{bailliage04} is
somewhat different.)

The \ProPub framework of \textbf{Dey, Zinn and Lud\"ascher}
\cite{dey11ssdbm}, based on Datalog, provides what the authors refer to
as ``policy-aware'' provenance sanitization. A provenance query is
expressed as a set of Datalog facts, asserting that the provenance for
certain data items is to be disclosed, plus additional requirements
relating to sanitization and disclosure. \ProPub works directly with a
provenance graph, which may not have been derived from an underlying
workflow. A sanitization requirement might assert that certain data
associated with a particular node is to be erased, that several nodes
are to be abstracted into a single node, or that some nodes are to be
deleted; a disclosure constraint might insist that a specific node is
always retained in the view. In addition there will usually be global
policies which hold across all queries (for example to outlaw ``false
dependencies'' of the kind illustrated earlier in
\figref{overview:provenance-view}), as well as the usual well-formedness
conditions such as acyclicity or bipartiteness.

A unique feature of \ProPub is its ability to detect conflicts in the
sanitization and disclosure requirements and to assist with their
resolution. When conflicts arise, \ProPub uses a ranking scheme and
various auto-correction strategies to resolve them, with the user also
able to intervene to withdraw or modify a constraint in the light of the
conflicts. For example in \figref{cyclic-provenance}, adapted from
\cite{dey11ssdbm}, a na\"ive abstraction of three nodes into a single
node $c_4$ violates both acyclicity and bipartiteness:

\begin{SCfigure}%[H]
%\begin{center}
\includegraphics[scale=0.8]{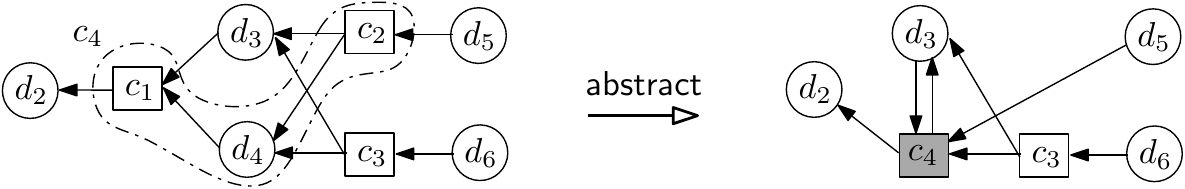}
%\end{center}
\caption{\ProPub: conflict detection}
\label{fig:cyclic-provenance}
\end{SCfigure}

\noindent In this case a possible resolution would be to include $d_3$
into the abstraction as well, removing the cycle and restoring
bipartiteness. Should applying a correction induce other conflicts, the
process of conflict resolution continues. Only when a conflict-free
variant of the query is obtained can a final sanitized view be derived.
Any constraints rescinded during conflict resolution are reported
alongside the sanitized view, providing a certain level of
``meta-provenance'', also a unique feature amongst the systems
considered here. For example, it might record that a spurious dependency
was tolerated in order to accommodate an abstraction. \ProPub's logical
foundation also means that the final view is guaranteed to have the
chosen disclosure and security properties.

\textbf{Davidson \etal}~\cite{davidson11pods,davidson13icdt} tackle a
rather different problem with \textbf{provenance views}. Workflows are
modelled as directed acyclic multigraphs (graphs with potentially more
than one edge between any two nodes). Edges are labeled with identifiers
called \emph{attributes} which identify the port that the edge starts
from; because workflows are acyclic, the semantics of a workflow can be
given as a relation $R$ over the set of all attributes, where each tuple
consists of the data values that arise during a possible execution.
(Equivalently, one can consider each tuple to be a labeling function
assigning data values to ports.) In \figref{survey:davidson:workflow}
below, adapted from \cite{davidson11pods}, the workflow consists of
three modules computing Boolean functions. Port $a_4$ of $m_1$ is
consumed by both $m_2$ and $m_3$.
The relation $R$ for this particular workflow in shown in the middle
of Figure\ref{fig:survey:davidson:workflow}.
Effectively $R$ is the natural join $R_1 \bowtie R_2 \bowtie R_3$ of the
relations $R_1$, $R_2$ and $R_3$ capturing the \emph{extension}
(input-output mapping) of the modules individually.

\begin{figure}
\begin{center}
\includegraphics[scale=0.8]{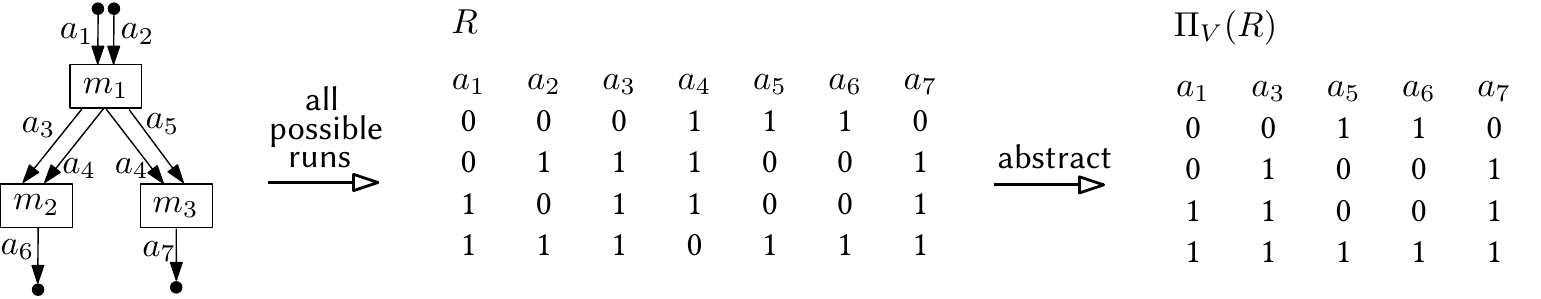}
\end{center}
\caption{Provenance views: hiding functional behaviour}
\label{fig:survey:davidson:workflow}
\end{figure}

Rather than hiding or abstracting parts of a particular run, Davidson
\etal are interested in hiding the extension of a sensitive module
$m_i$, namely the relation $R_i$, regardless of how many different
executions the user observes. They classify modules as either public,
whose behaviour is known \emph{a priori}, or private, whose behaviour
must be inferred by observing $R$. Their approach, which is
quantitative, is based on an extension of $\ell$-diversity
\cite{machanavajjhala07tkdd} which they call \emph{$\Gamma$-privacy}. A
view is specified by giving a set $V$ of visible attributes. The
relation $\Pi_V(R)$, the projection of $R$ to $V$
(\figref{survey:davidson:workflow}, right), defines the information that
is publicly visible through $V$. For any positive natural number
$\Gamma$, a private module is \emph{$\Gamma$-private} with respect to
$V$ if for each input, the number of possible outputs from that module
consistent with $\Pi_V(R)$ is greater than $\Gamma$. With only this
information, an attacker is unable to predict the output of the module
for a given input with probability greater than $1 / \Gamma$.

The first paper~\cite{davidson11pods} studies some specific cases,
including standalone private modules, multiple private modules, and
heterogeneous workflows with a mixture of private and public modules
where public modules can be ``privatized'' by renaming, so that their
functional behaviour is no longer known. They show that standalone
$\Gamma$-privacy is composable in a workflow consisting only of
private modules. The authors also study the problem of finding
minimum-cost views, given a cost function stating the penalty of being
denied access to hidden attributes. The second
paper~\cite{davidson13icdt} studies a more general solution for
heterogeneous workflows, which involves \emph{propagating} hiding,
i.e.~hiding attributes of public modules if they might disclose
information about hidden attributes of private modules. They present a
composability result generalizing the one for the all-private setting,
to \emph{single-predecessor} (that is, tree-like) workflows.

The privacy problem studied by Davidson \etal is interesting, but their
work so far has a number of drawbacks. In particular, the PTIME bounds
for the algorithms for mixed workflows~\cite{davidson13icdt} assume a
fixed domain size, which in turn means that the size of relation $R$ is
treated as a constant. If we take the domain size $d$ and number of
attributes $a$ into account, then the size of $R$ is $O(d^a)$, so
treating it as a constant may not be realistic. Moreover, it is also not
always clear how to choose sensible values of $\Gamma$. For example,
with a domain of $1024{\times}1024$, 8-bit grayscale images, $\Gamma$
may need to be much higher than $10^6$ to provide meaningful privacy,
because changing a single grayscale pixel does not hide much information.
(This criticism also pertains to other possibilistic definitions of
security properties, such opacity \cite{bailliage04} and
obfuscation~\cite{cheney11csf}.) Techniques from quantitative
information flow security \cite{clark02qapl}, quantitative
opacity~\cite{bryans13} or differential privacy~\cite{dwork06icalp} may
be relevant here.

The \textbf{provenance abstraction} approach of \textbf{Missier, Bryans,
  Gamble, Curcin and Danger} \cite{missier13}, implemented as \ProvAbs,
is based on graph quotienting and finding partionings that satisfy both
security needs and well-formedness constraints. Their provenance graphs
follow the \PROV model \cite{moreau13wc3} and its associated constraints
specification \cite{cheney13w3c}. First, Missier \etal consider simple
bipartite provenance graphs with node types representing activities and
entities, and define three basic graph operations \textsf{pclose},
\textsf{extend} and \textsf{replace}. Intuitively, \textsf{pclose} takes
a subgraph which is a candidate for replacement, and grows it until it
is convex (there are no paths that lead out of the subgraph and back in
again); \textsf{extend} further grows the subgraph until both its
``input'' nodes and its ``output'' nodes are homogeneous with respect to
node type; and \textsf{replace} contracts such subgraphs to single nodes
and adjusts edges to preserve paths.

\begin{figure}[h]
\begin{center}
\includegraphics[scale=0.8]{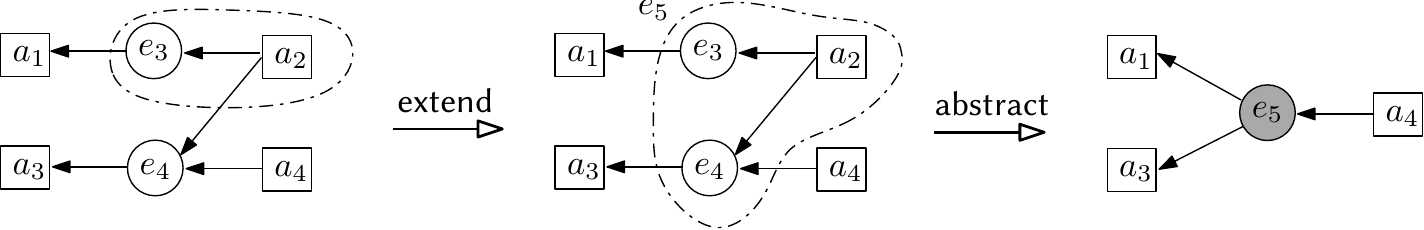}
\end{center}
\caption{\ProvAbs: growing a partition so that abstraction preserves bipartiteness}
\label{fig:survey:missier}
\end{figure}

\figref{survey:missier}, adapted from \cite{missier13}, illustrates
\textsf{extend} and \textsf{replace}. In \figsubref{survey:missier}{a},
the user selects activity $a_2$ and entity $e_3$ for abstraction.
Replacing these two nodes by either an activity or an entity whilst
preserving paths would violate bipartiteness. In
\figsubref{survey:missier}{b}, \textsf{extend} is used to grow the
target subgraph to include $e_4$, so that the output nodes of the target
subgraph are uniformly entities. Replacing the subgraph by a single
abstract entity $e_5$ in \figsubref{survey:missier}{c} is now valid,
although it coarsens the (transitive) dependencies by introducing a path
between $a_4$ and $a_1$.

Having shown how these transformations can be used to preserve basic
validity constraints, Missier \etal go on to consider graphs which
incorporate the \PROV \textsf{agent} node type and associated relations
such as attribution and delegation. They consider three cases of
increasing sophistication. Grouping a homogeneous set of agents into a
single abstract agent is relatively straightforward. Grouping agents and
entities together is trickier; the type of the target abstract node
(entity or agent) must be specified, and in order to maintain the
type-correctness of certain relations between actions and agents
(\textsf{waw}, ``was associated with'') and between entities and agents
(\textsf{wat}, ``was attributed to''), the subgraph to be abstracted
must made larger. Finally, grouping arbitrary node types together
presents the additional difficulty of agent-to-agent delegation edges
(\textsf{abo}, ``acted on behalf of''), which require similar treatment.

Like \ProPub, a key feature of \ProvAbs is that transformations operate
directly on the provenance graph, and are thus more suited to situations
where there is no underlying workflow. Missier \etal claim that their
system avoids introducing spurious dependencies between nodes. However,
their views are quotients, which in general over-approximate
dependencies, so technically this claim is only correct for provenance
applications where dependency is not required to be transitive.

The work of \textbf{Cadenhead, Khadilkar, Kantarcioglu and
  Thuraisingham} \cite{cadenhead11} on \textbf{provenance redaction} is
also based on graph rewriting. Their provenance graphs are tripartite
and conform to the Open Provenance Model's labeled DAG
format~\cite{opm11}. ``Redacting'', or sanitizing, such a graph has two
phases. First, the sensitive region $G_Q$ (typically a single node or a
path between two nodes) of the original graph $G$ is isolated using a
graph query $Q$. Then, this region of the graph is transformed according
to an obfuscation policy expressed as rewrite rules. A rewrite rule has
two components: a production rule $r : L \rightarrow R$, where $L$ is
matched against subgraphs of $G_Q$, plus an \emph{embedding}
specification, which determines how edges are to be connected to $R$
once it has replaced $L$. The rewrites involve graph operations such as
vertex contraction, edge contraction, path contraction and node
relabeling.

\begin{SCfigure}
\includegraphics[scale=0.8]{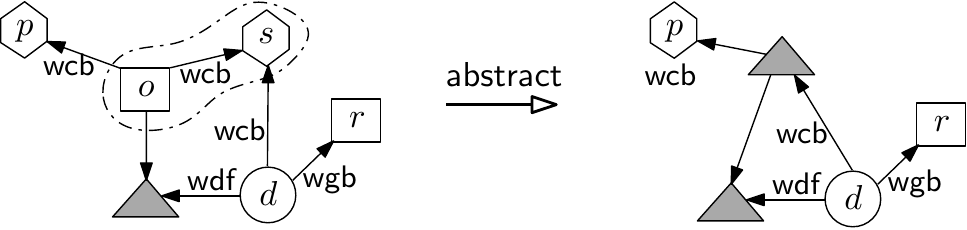}
\caption{Provenance redaction: abstraction by edge contraction}
\label{fig:survey:cadenhead}
\end{SCfigure}

In \figref{survey:cadenhead} above, adapted from \cite{cadenhead11},
hexagons represent agents, rectangles represent processes, and circles
represent artifacts. In the left graph, the gray triangle indicates an
area of the graph that was previously redacted. On the right, a further
subgraph is redacted by contracting the the \textsf{wcb} (``was
controlled by'') edge relating a heart operation $o$ to the surgeon $s$
who carried out the operation, and replacing the two nodes by another
gray triangle.

Cadenhead \etal's work is implementation-focused. Several formal
definitions are given but not always made use of, and neither are
their provenance or disclosure properties analyzed. One issue they do
not appear to address, in contrast for example to Missier \etal
(above), is preservation of basic well-formedness properties of the
provenance graph. While edge contraction (as a particular kind of
quotient) preserves dependencies, the interaction with tripartiness is
potentially problematic. For example in the view in
\figref{survey:cadenhead}, the new triangle has both an incoming
\emph{and} an outgoing \textsf{wcb} edge, because it subsumes both an
agent and a process. Moreover, as the authors themselves point out,
the obfuscation policy is only applied to a subgraph $G_Q$ of the
original graph $G$. Sensitive information available elsewhere in $G$
will not be subject to the policy.  Information flow techniques
\cite{denning77cacm} may be relevant here.

\section{Conclusions and future directions}

We conclude our survey with a brief feature comparison, summarised in
\tableref{feature-comparison}. The column headings refer to broad
feature areas (discussed in more detail below); $\FULL$ indicates
reasonably comprehensive support for that feature, $\NONE$ little or no
support, and $\HALF$ somewhere in between. Necessarily this is a
somewhat simplistic assessment.

\newcommand{\colWidth}{1cm}
\begin{table}[H]
\small
  \begin{center}
    \begin{tabular}{|l|c|c|c|c|c|c|c|c|}
%   \begin{tabular}{|l|p{\colWidth}|p{\colWidth}|p{\colWidth}|p{\colWidth}|p{\colWidth}|p{\colWidth}|p{\colWidth}|}
\hline
System &
\featIntegrity &
\featDependency &
\featAccess &
\featQuery &
\featSemantic &
\featFormal &
\featConflict &
\featMeta
\\
\hline
\Zoom~\cite{biton08icde,cohen-boulakia08}             & \HALF & \FULL & \HALF & \NONE & \HALF & \NONE & \NONE & \NONE \\
Security views~\cite{chebotko08}                      & \HALF & \HALF & \FULL & \NONE & \NONE & \NONE & \HALF & \NONE \\
Surrogates~\cite{blaustein11pvldb}                    & \NONE & \HALF & \FULL & \NONE & \NONE & \NONE & \HALF & \NONE \\
\ProPub~\cite{dey11ssdbm}                             & \HALF & \HALF & \FULL & \FULL & \NONE & \FULL & \FULL & \HALF \\
Provenance views~\cite{davidson11pods,davidson13icdt} & \NONE & \HALF & \HALF & \HALF & \HALF & \FULL & \NONE & \NONE \\
\ProvAbs~\cite{missier13}                             & \FULL & \FULL & \FULL & \HALF & \NONE & \NONE & \HALF & \NONE \\
Provenance redaction~\cite{cadenhead11}               & \HALF & \FULL & \FULL & \FULL & \NONE & \NONE & \NONE & \NONE \\
\hline
    \end{tabular}
  \end{center}
\caption{Feature comparison for the approaches surveyed}
\label{table:feature-comparison}
\end{table}
\vspace{-25pt} % stop LaTeX putting a big gap here

\paragraph{Integrity.}

We divide integrity features into basic integrity maintenance
(\textbf{\featIntegrity}) and integrity of causal or dependency
structure (\textbf{\featDependency}). Even systems that make some effort
to preserve the latter, such as provenance redaction, may in so doing
violate low-level integrity constraints. In the future it seems likely
that users will take low-level integrity for granted.

Preservation or reflection of dependency structure is more challenging
because of the inherent tensions with obfuscation requirements. When
arbitrary nodes or edges can be deleted, then the user may be
responsible for repairing the damage, as with security views or
surrogates. \ProPub offers greater automation through conflict
detection; \ProvAbs and provenance redaction make safer (if simplistic)
assumptions, by working mainly with quotient views.

\paragraph{Sanitization.}

Sanitization features range from explicit fine-grained access control
(\textbf{\featAccess}), which all systems provide in some form or
another, to query-based abstraction (\textbf{\featQuery}), as offered by
\ProPub and provenance redaction. Query-based systems typically subsume
fine-grained access control, via fine-grained queries.

\paragraph{Formal and semantic properties.}

Few of the surveyed systems consider the problem of relating provenance
views to the semantics of the underlying system
(\textbf{\featSemantic}). Instead, they operate directly on provenance
graphs, without regard to how the graph was created. This is flexible,
but means one cannot easily treat the provenance view as an (abstracted)
account of how something was computed. \Zoom stands out in this respect,
in relating provenance views to workflow views for simple kinds of
workflow. On the other hand, this is a hard problem to solve in a
general way.

Few existing systems provide formal guarantees of obfuscation or
disclosure properties (\textbf{\featFormal}). \ProPub has the advantage
of a solid logical foundation. The $\Gamma$-privacy of provenance views
is a formal notion of (quantitative) opacity, but the goal is somewhat
different from the other systems considered.

\paragraph{Conflict detection and resolution.}

As mentioned, \ProPub stands out in being able to automatically detect
conflicts between obfuscation and disclosure requirements
(\textbf{\featConflict}), thanks to its logic-based approach. It is also
the only system which makes conflict resolution an explicit and
persistent part of the process, providing a certain level of
``meta-provenance'' for the sanitization process (\textbf{\featMeta}).
If provenance security techniques are widely adopted, it seems likely
that how provenance is manipulated to hide or reveal information will
itself often be the point of interest (cf.~``provenance of provenance''
\cite{moreau13wc3}).

\vspace{5pt}
\noindent Undoubtedly, controlling access to sensitive provenance
metadata is of growing importance, and moreover we sometimes simply want
to deliver provenance information at a particular level of detail.
However, as the summary above highlights, current methods for provenance
sanitization are immature. Future effort should focus on semantics,
formal guarantees, and techniques for detecting and resolving
conflicting policies.

\paragraph*{Acknowledgments}
We are grateful to Jeremy Bryans, Brian Gamble, and Paolo Missier for
comments on this paper.  Effort sponsored by the Air Force Office of
Scientific Research, Air Force Material Command, USAF, under grant
number FA8655-13-1-3006. The U.S. Government and University of
Edinburgh are authorized to reproduce and distribute reprints for
their purposes notwithstanding any copyright notation thereon.

\bibliographystyle{abbrv}
\bibliography{paper}

\begin{thebibliography}{10}

\bibitem{bailliage04}
R.~D. Bailliage and L.~Mazar\'e.
\newblock Using unification for opacity properties.
\newblock In {\em Proceedings of WITS 2004}, pages 165--176, 2004.

\bibitem{biton08icde}
O.~Biton, S.~Cohen-Boulakia, S.~B. Davidson, and C.~S. Hara.
\newblock Querying and managing provenance through user views in scientific
  workflows.
\newblock In {\em ICDE}, pages 1072--1081. IEEE, 2008.

\bibitem{blaustein11pvldb}
B.~T. Blaustein, A.~Chapman, L.~Seligman, M.~D. Allen, and A.~Rosenthal.
\newblock Surrogate parenthood: Protected and informative graphs.
\newblock {\em PVLDB}, 4(8):518--527, 2011.

\bibitem{braun08hotsec}
U.~Braun, A.~Shinnar, and M.~Seltzer.
\newblock Securing provenance.
\newblock In {\em Proceedings of the 3rd conference on Hot topics in security},
  pages 4:1--4:5, 2008.

\bibitem{bryans13}
J.~Bryans, M.~Koutny, and C.~Mu.
\newblock Towards quantitative analysis of opacity.
\newblock In C.~Palamidessi and M.~Ryan, editors, {\em Trustworthy Global
  Computing}, volume 8191 of {\em Lecture Notes in Computer Science}, pages
  145--163. Springer, 2013.

\bibitem{buneman06sigmod}
P.~Buneman, A.~P. Chapman, and J.~Cheney.
\newblock Provenance management in curated databases.
\newblock In {\em SIGMOD 2006}, pages 539--550, 2006.

\bibitem{cadenhead11}
T.~Cadenhead, V.~Khadilkar, M.~Kantarcioglu, and B.~Thuraisingham.
\newblock Transforming provenance using redaction.
\newblock In {\em SACMAT}, pages 93--102, New York, NY, USA, 2011. ACM.

\bibitem{chebotko08}
A.~Chebotko, S.~Chang, S.~Lu, F.~Fotouhi, and P.~Yang.
\newblock Scientific workflow provenance querying with security views.
\newblock In {\em WAIM 2008}, pages 349--356, 2008.

\bibitem{cheney11csf}
J.~Cheney.
\newblock A formal framework for provenance security.
\newblock In {\em CSF}, pages 281--293. IEEE, 2011.

\bibitem{cheney13w3c}
J.~Cheney, P.~Missier, L.~{Moreau (eds.)}, and T.~{De Nies}.
\newblock Constraints of the {PROV} data model.
\newblock {W3C} recommendation, W3C, April 2013.

\bibitem{chong09tapp}
S.~Chong.
\newblock Towards semantics for provenance security.
\newblock In J.~Cheney, editor, {\em TaPP 2009}. USENIX, 2009.

\bibitem{clark02qapl}
D.~Clark, S.~Hunt, and P.~Malacaria.
\newblock Quantitative analysis of the leakage of confidential data.
\newblock {\em Electronic Notes in Theoretical Computer Science}, 59(3):238 --
  251, 2002.
\newblock QAPL 2001.

\bibitem{cohen-boulakia08}
S.~Cohen-Boulakia, O.~Biton, S.~Cohen, and S.~Davidson.
\newblock Addressing the provenance challenge using zoom.
\newblock {\em Concurrency and Computation: Practice and Experience},
  20(5):497--506, Apr. 2008.

\bibitem{davidson08icmd}
S.~B. Davidson and J.~Freire.
\newblock Provenance and scientific workflows: Challenges and opportunities.
\newblock In {\em Proceedings of SIGMOD 2008}, pages 1345--1350, New York,
  2008. ACM.

\bibitem{davidson11pods}
S.~B. Davidson, S.~Khanna, T.~Milo, D.~Panigrahi, and S.~Roy.
\newblock Provenance views for module privacy.
\newblock In {\em PODS}, pages 175--186, 2011.

\bibitem{davidson13icdt}
S.~B. Davidson, T.~Milo, and S.~Roy.
\newblock A propagation model for provenance views of public/private workflows.
\newblock In {\em ICDT}, pages 165--176, New York, NY, USA, 2013. ACM.

\bibitem{denning77cacm}
D.~E. Denning and P.~J. Denning.
\newblock Certification of programs for secure information flow.
\newblock {\em Communications of the ACM}, 20(7):504--513, July 1977.

\bibitem{dey11ssdbm}
S.~C. Dey, D.~Zinn, and B.~Lud{\"a}scher.
\newblock {ProPub}: Towards a declarative approach for publishing customized,
  policy-aware provenance.
\newblock In {\em SSDBM}, pages 225--243, 2011.

\bibitem{dwork06icalp}
C.~Dwork.
\newblock Differential privacy.
\newblock In {\em ICALP}, pages 1--12. Springer, 2006.

\bibitem{hasan07sss}
R.~Hasan, R.~Sion, and M.~Winslett.
\newblock Introducing secure provenance: problems and challenges.
\newblock In {\em Proceedings of StorageSS 2007}, pages 13--18, New York, NY,
  USA, 2007. ACM.

\bibitem{hasan09tos}
R.~Hasan, R.~Sion, and M.~Winslett.
\newblock Preventing history forgery with secure provenance.
\newblock {\em Trans. Storage}, 5:12:1--12:43, December 2009.

\bibitem{lu13vldbj}
W.~Lu, G.~Miklau, and N.~Immerman.
\newblock Auditing a database under retention policies.
\newblock {\em VLDB Journal}, 22(2):203--228, 2013.

\bibitem{lyle10tapp}
J.~Lyle and A.~Martin.
\newblock Trusted computing and provenance: better together.
\newblock In {\em Proceedings of TAPP 2010}, Berkeley, CA, USA, 2010. USENIX
  Association.

\bibitem{machanavajjhala07tkdd}
A.~Machanavajjhala, D.~Kifer, J.~Gehrke, and M.~Venkitasubramaniam.
\newblock L-diversity: Privacy beyond k-anonymity.
\newblock {\em ACM Trans. Knowl. Discov. Data}, 1(1), Mar. 2007.

\bibitem{martin12tapp}
A.~Martin, J.~Lyle, and C.~Namilkuo.
\newblock Provenance as a security control.
\newblock In {\em Proceedings of TaPP 2012}, pages 3--3, Berkeley, CA, USA,
  2012. USENIX Association.

\bibitem{missier13}
P.~Missier, J.~Bryans, C.~Gamble, V.~Curcin, and R.~Danger.
\newblock Provenance graph abstraction by node grouping.
\newblock Technical Report CS-TR-1393, Newcastle University, 2013.

\bibitem{moreau10ftws}
L.~Moreau.
\newblock The foundations for provenance on the web.
\newblock {\em Foundations and Trends in Web Science}, 2(2--3), 2010.

\bibitem{opm11}
L.~Moreau, B.~Clifford, J.~Freire, J.~Futrelle, Y.~Gil, P.~Groth,
  N.~Kwasnikowska, S.~Miles, P.~Missier, J.~Myers, B.~Plale, Y.~Simmhan,
  E.~Stephan, and J.~Van~den Bussche.
\newblock The {OPM} core specification (v1.1).
\newblock {\em Future Generation Computer Systems}, 27(6):743--756, June 2011.

\bibitem{moreau13wc3}
L.~Moreau and P.~{Missier (eds.)}.
\newblock {PROV-DM: The PROV Data Model}.
\newblock {W3C Recommendation} REC-prov-dm-20130430, 2013.

\bibitem{zhang09sdm}
J.~Zhang, A.~Chapman, and K.~Lefevre.
\newblock Do you know where your data's been? --- tamper-evident database
  provenance.
\newblock In {\em Proceedings of SDM 2010}. Springer-Verlag, 2009.

\end{thebibliography}

\end{document}